# A High Throughput Study of both Compositionally Graded and Homogeneous Fe-Pt Thin Films

Yuan Hong[1], Isabelle de Moraes[1], Gabriel Gomez Eslava[1], Stéphane Grenier[1], Edith Bellet-Amalric[2], Andre Dias[1], Marlio Bonfim[3], Laurent Ranno[1], Thibaut Devillers[1] and Nora M. Dempsey[1]

[1]*Univ. Grenoble Alpes, CNRS, Grenoble INP, Institut NEEL, 38000 Grenoble, France*
[2]*Univ. Grenoble Alpes, CEA, Grenoble INP, IRIG, PHELIQS, 38000 Grenoble, France*
[3]DELT, *Universidade Federal do Paraná, Curitiba, Brazil*

Abstract

Compositionally graded Fe-Pt thin films were prepared on stationary 100 mm Si substrates by magnetron sputtering a base target of Fe on which a piece of Pt is asymmetrically positioned. Energy Dispersive X-Ray analysis was used to map the variation in film composition across the substrate, as a function of the size of the Pt piece. A scanning polar Magneto-Optical-Kerr-Effect system was used to probe the influence of composition and post-deposition annealing conditions (temperature and time) on coercivity. In this way the maximum coercivity achievable for the sputtering system used could be established in a high throughput fashion. The evolution in coercivity with composition was correlated with the formation of $L1_0$ FePt and changes in its lattice parameters, as determined by scanning X-ray diffraction. High throughput coercivity mapping was then carried out on homogeneous Fe-Pt thin films of different composition treated to different annealing conditions. This study serves as a step towards the integration of coercive FePt films into collectively fabricated devices.

1. Introduction

The functional magnetic properties of magnetic films depend on their crystallographic structure and microstructure (grain size, grain orientation, secondary and grain boundary phases…). These in turn depend on the film's chemical composition as well as the deposition and annealing conditions applied during film fabrication. Combinatorial thin film studies, based on the preparation and characterisation of compositionally graded films, are being used for the high-throughput screening and optimization of a range of functional materials [1-4], including hard magnetic films [5, 6]. In a recent perspective, A. Ludwig discussed the possibilities offered by combinatorial synthesis and high-throughput characterization in combination with computational methods to efficiently identify new materials in multi-dimensional search spaces, including ternary, quaternary and pentanary compounds [7]. He also commented that although combinatorial deposition methods for thin films and multilayers are well-established, further automatization of high throughput characterization methods is needed to accelerate materials discovery.

The $L1_0$ FePt phase of the Fe-Pt system has high magnetocrystalline anisotropy ($K_1$ = 6 MJ/m$^3$) and excellent chemical stability [8] and it is studied for use in ultrahigh-density storage media and magnetic micro/nano-systems. Ludwig et al. prepared Fe-Pt wedged multilayer thin films with a broad composition range and studied how coercivity and magnetisation varied as a function of composition and annealing conditions [5]. The advantage of this combinatorial approach is that all compositions are treated to exactly the same set of annealing conditions, thus avoiding sample-to-sample variations which may occur when samples are prepared and annealed individually. The continuous composition gradient also allows to accurately determine the composition at which a transition occurs. However, the magnetic characterization step used was a bottleneck as it required samples to be diced for individual measurement in a high field magnetometer, and such measurements typically take some tens of minutes to perform.

Comparison of the results from that study and other literature shows that the optimum reported composition and annealing conditions for maximizing coercivity or anisotropy energy in $L1_0$ FePt based bulk and thin film samples vary from publication to publication [9-15]. In the case of composition, the reported optimum values varied by a few atomic percent and moves from the Fe-rich side [5, 15, 16] to the Pt-rich side [14] of the phase diagram. Thus, when one wants to fabricate coercive FePt films in a given deposition system for a given application, one is faced with the task of selecting the appropriate film composition and applying appropriate annealing conditions. In this work, we prepared compositionally graded Fe-Pt films and characterised them both magnetically and structurally in a high throughput fashion, so as to establish near-optimum composition and annealing conditions. We then used high throughput magnetic characterisation to optimise the composition and annealing conditions of homogeneous films, so that they can be used to fabricate high coercivity films over large surfaces, rendering them suitable for integration into collectively fabricated devices.

2. Experimental

Fe-Pt films were fabricated by magnetron sputtering (DP850, Alliance Concept) of an Fe target (99.9%) of diameter 75 mm, partially covered by Pt foil (99.95%), onto stationary substrates. In the main study, compositionally graded films were produced by asymmetrically positioning the pieces of Pt foil on the Fe target and the composition gradient was varied by changing the size of the Pt pieces: 25 mm × 25 mm (target #1), 37.5 mm × 25 mm (target #2) and 50 mm × 25 mm (target #3) (Figure 1). In a secondary study, compositionally homogeneous films were produced by symmetrically positioning pieces of Pt foil on the concentric track of the Fe base target where the sputtering rate is highest, and the composition was varied by changing the size and disposition of the Pt pieces (Figure 2). Films were deposited at room temperature with a bias voltage of 100 V, sputtering power of 40 W and an argon flow of 20 sccm on thermally oxidised Si substrates of diameter 100 mm. All films had the following architecture: Si/SiO$_2$ (100 nm) / Ta (10 nm) / Fe-Pt (x nm) / Ta (10 nm), with nominal thickness of x = 100 nm for the compositionally graded films and x = 50 nm for the homogeneous films. TEM imaging of select pieces from an FePt film together with AFM measurements of patterned metallic films made with the same sputtering system reveal that the film thickness is maximum at the centre and drops by roughly 30% close to the wafer edge. 1D line-scans or 2D maps of composition of as-deposited films were made using Energy Dispersive X-Ray (EDX, Oxford Instruments) analysis (spot size ~ 1 μm) in a scanning electron microscope (SEM, ZEISS). The nominal accuracy of the estimated composition values is of the order of ±1-2 at.%. Post-deposition annealing was carried out on full substrates using a rapid thermal annealing furnace (RTA, Jipelec), while it was carried out on quarter wafers using a tube furnace. High throughput magnetic characterisation was performed using an in-house developed scanning polar Magneto-Optic Kerr effect (MOKE) system with an integrated coolant-free bi-polar pulsed magnetic field generator and a laser spot size of roughly 50 μm [17]. The maximum field strength applied at the film surface during a given loop measurement was 4 T, which is significantly higher than the maximum field values of 0.3 T [18] and 2.3 T [19] used in previous high throughput MOKE studies of hard magnetic films. The duration of individual field pulses is of the order of 16 μs, and the delay between positive and negative field pulses is roughly 10 ms. The coercivity of some reference sample pieces extracted from a compositionally graded film was measured using a SQUID-VSM (MPMS-3, Quantum Design). The evolution of crystal structure along the composition gradient was characterized in a high throughput fashion using scanning X-ray diffraction (XRD, Rigaku SmartLab) with Cu-K$_\alpha$ radiation.

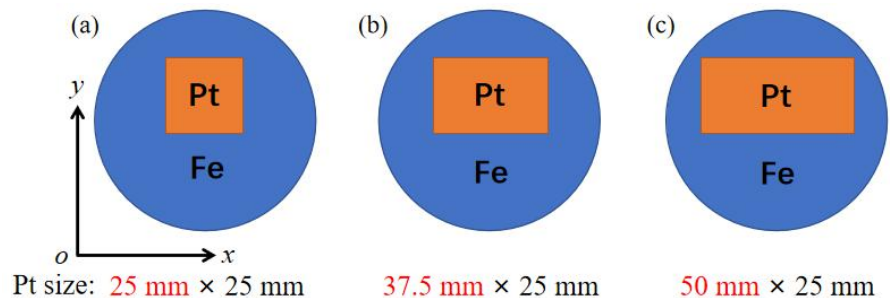

Fig. 1 Schematic diagrams of the targets used to produce compositionally graded Fe-Pt films indicating the position and size of Pt foil on a base Fe target: (a) target #1, (b) target #2, (c) target #3

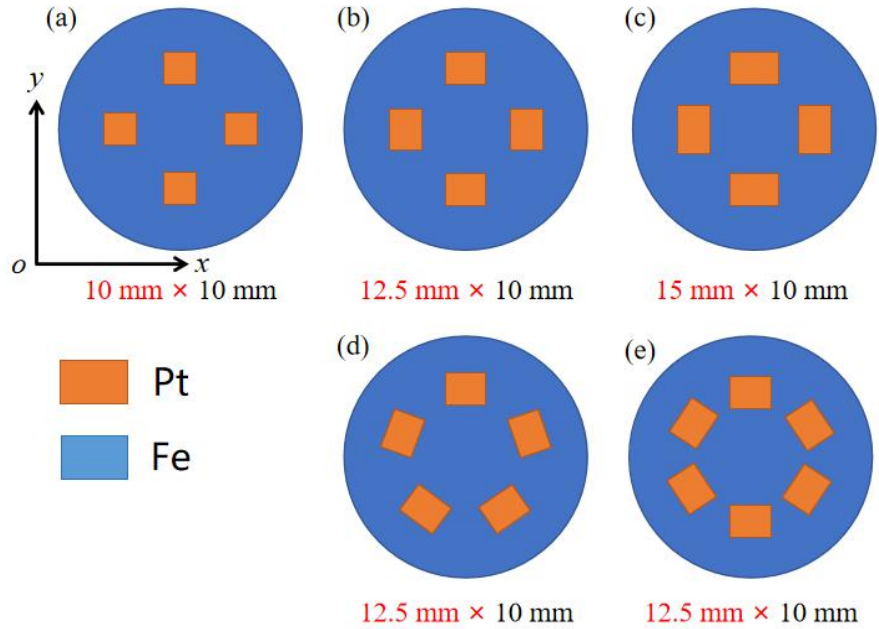

Fig. 2 Schematic diagrams of the targets used to produce compositionally homogeneous FePt films indicating the position and size of Pt foils on a base Fe target: (a) target #4, (b) target #5, (c) target #6, (d) target #7, (e) target #8

3. Results and discussion

*3.1 Compositionally graded films*

1D composition line-scans measured along the horizontal bi-sector (i.e., parallel to the *y* direction) of films made from targets #1 - #3 are compared in Figure 3. The equiatomic composition of $Fe_{50}Pt_{50}$ is found close to the centre of the film made with target #2, while it is positioned on the edges of the films made with the other two targets. 2D EDX scans consisting of a matrix of 19×17 points (distance between each point = 5 mm, measuring time per point = 2 minutes) were made to probe for composition variations in both the *x* and *y* directions. Such a measurement takes roughly 12 hours to perform. 2D composition maps of the films made using targets #2 and # 3 are compared in Figure 4. The same colour scale bar is used in both figures, to emphasize the difference in composition spread. The observed variations in composition in the *x*-direction, which gives a somewhat wavy shape to the iso-composition lines, is attributed to slightly asymmetric positioning of the Pt pieces on the Fe base target and/or to inhomogeneities of the plasma inside the sputtering machine. Such variations are not picked up in the 1D vertical scans, thus the interest in 2D scans. Having the expected optimum composition for the formation of $L1_0$ FePt close to the centre of the film allows the study of structural and magnetic properties of films deposited under exactly the same conditions, in a composition range which is sure to include the optimum composition for a given annealing treatment. We thus identified target #2 as the most suitable target and all results reported on compositionally graded films hereafter concern films made from this target. The atomic composition of each element spans roughly 33-67% across the 100 mm substrate, which is a little narrower than the composition span of 25-75% reported by Ludwig et al. for compositionally graded Fe-Pt films deposited on 150 mm substrates [5]. A difference in the shape of the composition variations achieved (non-linear here vs. linear in the study of Ludwig et al.) may be attributed to differences in the sputtering systems and the fabrication protocols used (Ludwig et al. deposited alternating opposing wedges of Fe and Pt).

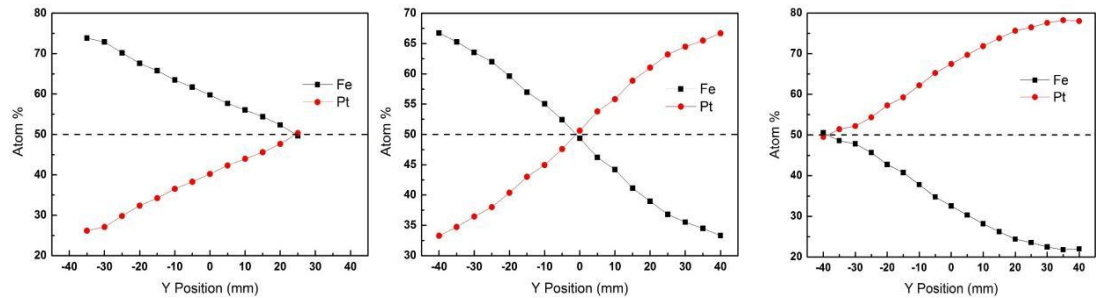

Fig. 3 1D EDX line scans of Fe and Pt content measured along the horizontal bi-sector (i.e., parallel to the *y*-axis) of samples fabricated using targets (a) #1 (a), (b) #2 and (c) #3.

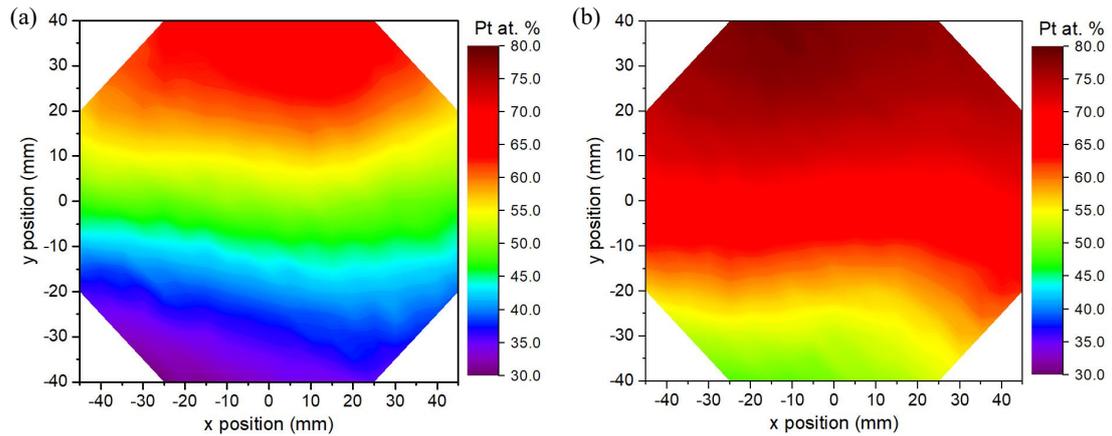

Fig. 4 2D EDX maps of Pt content measured over full wafers of films made using (a) target # 2 and (b) target # 3.

A set of compositionally graded Fe-Pt films prepared under identical conditions using target #2 were then annealed under different conditions and scanning high field MOKE was used for high throughput magnetic characterization. A first set of samples were annealed for 30 min at 500°C, 600°C and 700°C. Maps of MOKE loops of these samples are compared in Figure 5. All three MOKE maps reveal the same overall trend, with coercivity increasing from practically zero in the Fe-rich bottom part of the wafer to a maximum value above the mid-point of the wafer followed by a reduction in the Pt-rich top of the wafer. Coercivity is a signature of the presence of the $L1_0$ FePt phase as neither $FePt_3$ nor $Fe_3Pt$ phases have significant values of magnetocrystalline anisotropy. The variations in coercivity values seen in the MOKE maps are attributed to a combination of a number of factors. Firstly, the fact that the optimum annealing conditions for the ordering of the $L1_0$ phase depend on composition, as clearly demonstrated in calorimetry studies of Fe-Pt films [10] and in-situ neutron diffraction studies of ball milled Fe-Pt [8]. Secondly, the fact that coercivity is an extrinsic property that depends on microstructure (grain size, defects...), which can be expected to vary with composition and annealing conditions, even in single phase samples. Thirdly, intimate mixing of the high anisotropy $L1_0$ phase with soft magnetic $FePt_3$ or $Fe_3Pt$ phases could lead to a drop in coercivity due to exchange and/or dipolar interactions between the phases. Finally, intimate mixing of the high anisotropy L1$_0$ phase with non-magnetic Fe-rich or Pt-rich phases could also modify coercivity through a modification in the microstructure.

The surface area over which relatively high coercivity values (> 1T) are found is somewhat larger in the film annealed at the intermediate temperature of 600°C, and the maximum value of coercivity measured in this film was roughly 1.6T. The highest values of coercivity (1.7-1.8 T) were found in the film annealed at 700°C (positions identified by the red boxes in Figure 5c). The fact that maximum coercivity is found on the edge of the film may be attributed to a drop in film thickness from the centre to the edge of the film. The estimated 30% drop in film thickness at the film edge is thought to be too small to modify L1$_0$ ordering kinetics [10]. However, it may be enough to lead to an increase coercivity through a reduction in grain size.

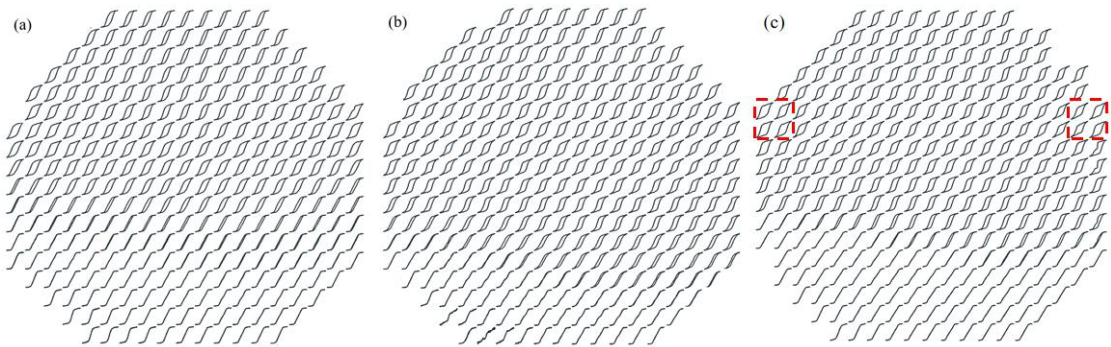

Fig. 5 2D Scanning polar MOKE maps of wafers, made from target #2, annealed at (a) 500°C, (b) 600°C and (c) 700°C for 30 minutes (maximum applied field = 4 T; MOKE intensity values are normalised).

A side-by-side comparison of the EDX map of an as-deposited film made using target #2, and a map of coercivity values extracted from the MOKE loops of the film annealed at 600°C for 30 minutes (shown in Figure 5b), is made in Figure 6. The coercivity is given by the averaged applied field value when the MOKE signal crosses through zero in the downward and upward branches of the loop. The horizontal variations in coercivity can clearly be associated with horizontal variations in composition. Coercivity is plotted as a function of Pt content, measured across the entire wafer, in Figure 6c. Coercivity is seen to emerge at around 37.5 at.% Pt, peaking at 1.6 T in the range 55-60 at.% Pt and then dropping to just under 1 T at the max Pt content (~ 67.5 at.%). The spread in coercivity values seen for a given composition, which reaches 10% of the peak value, may reflect local variations in microstructure and film thickness, as well as the error associated with coercivity estimation from MOKE loops. We recall here that the aim of the present study is to identify the optimum nominal composition and annealing conditions for a given sputtering system to produce coercive FePt for subsequent integration into devices, rather than a fundamental study of the influence of composition variations on the coercivity achieved.

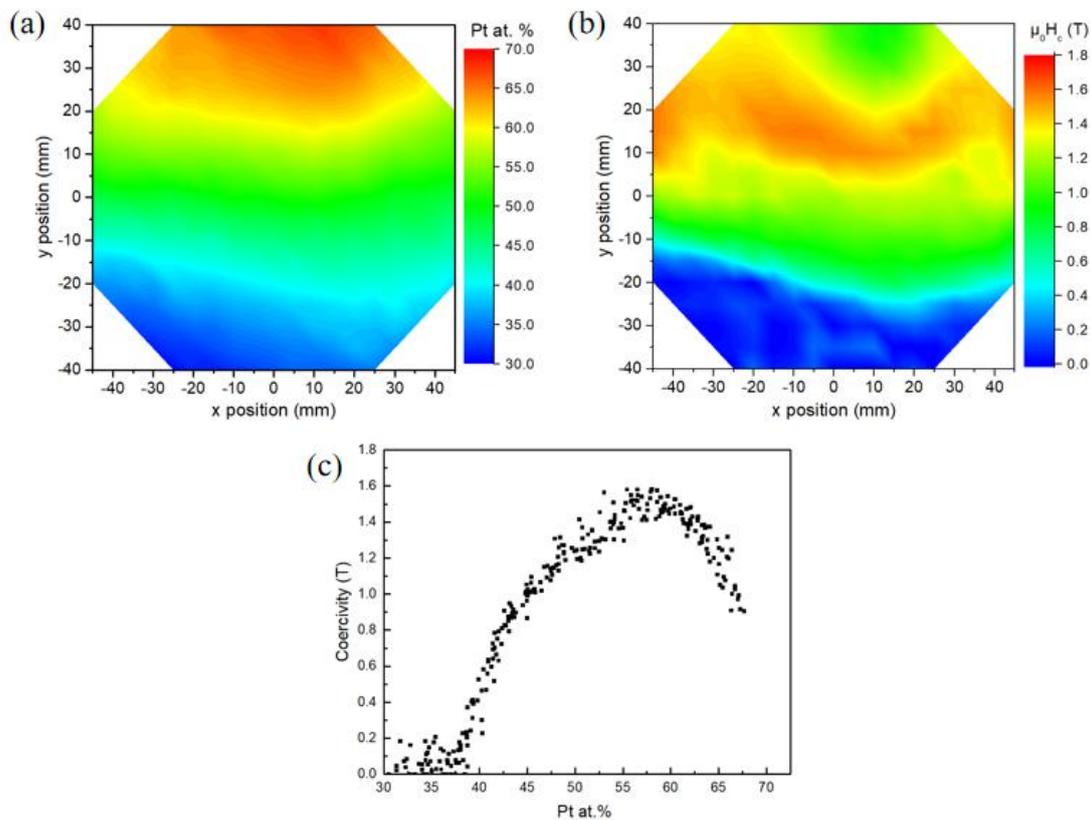

Fig. 6 Comparison of (a) the Pt content EDX map of an as-deposited film made using target #2, and (b) a coercivity map extracted from MOKE loops measured on a such a film annealed at 600°C for 30 minutes (see Figure 5b); (c) plot of coercivity as a function of Pt content, measured across the entire wafer.

Phase formation and the degree of ordering depends on both the temperature and time of annealing, and maps of MOKE loops of samples annealed at 600°C for durations of 1, 10 and 30 minutes are compared in Figure 7. The area of the zone showing high coercivity is found to increase with increasing annealing time.

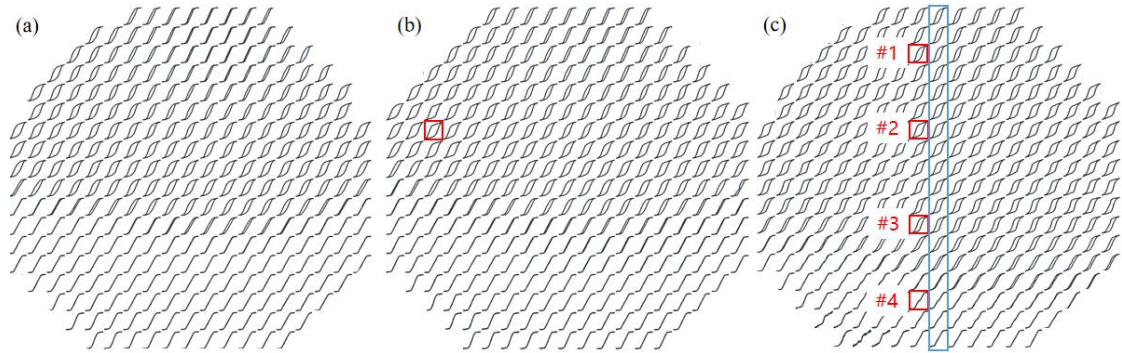

Fig. 7 2D Scanning polar MOKE maps of wafers, made from target #2, annealed at 600℃ for (a) 1 min, (b) 10 min (c) 30 min. Reference samples at the positions #1 to #4 were diced out of the wafer to make SQUID-VSM measurements.

To relate the oberserved changes in magnetic properties with variations in crystallographic structure, X-ray diffraction analysis was carried out. An XRD waterfall pattern measured along a vertical section of the film annealed at 600°C for 30 min., identified by a blue rectangle in Figure 7c, is shown in Figure 8. The distance between each measurement point is 2.5 mm while the X-ray beam spot is roughly 2-3 mm, allowing us to probe changes in crystallographic phases in an almost continuous fashion. The experimental data set is overlaid with the diffraction patterns of chemically ordered $L1_2$-$FePt_3$, $L1_0$-FePt and $L1_2$-$Fe_3Pt$, taken from ICSD reference cards 04-017-4976, 03-065-9121 and 04-001-3171, respectively. Peak indexation is the same for all three phases, with the $L1_0$ phase having additional peaks due to its tetragonal nature. The Fe-Pt diffraction peaks shift from higher to lower angle as we move up from the lower end of the wafer, indicating an expansion in lattice volume with increasing Pt content, since the metallic radius of Pt (138.5 pm) is much larger than that of Fe (126 pm). The experimental peaks at the Fe-rich end of the film are at a higher angle than those expected for ordered $Fe_3Pt$, as are the peaks of disordered

Fe$_3$Pt [20]. This, together with the fact that only fundamental peaks ((111), (200) and (202)) are clearly identified, indicates the presence of disordered and/or off-stoichiometric Fe$_{3+x}$Pt$_{1-x}$ in this region of the film. A striking feature of this graph is the splitting of certain peaks ((200)/(002), (220)/(202), (311)/(113)), starting at around -25 mm on the *y*-axis, which can be clearly associated with the formation of the tetragonally distorted L1$_0$ FePt phase, with the degree of distortion increasing with Pt content. At the Pt-rich end of the film, the 2θ positions of most reflections are closer to those of L1$_0$ FePt than FePt$_3$, though lower angle shoulders indicate the presence of a Pt-rich phase. This is clearly seen in Figure 9, which zooms in on the (111) diffraction peaks. The high relative intensity of the (111) and (222) peaks, compared to the expected intensities for an isotropic film (as indicated by the height of the lines indicating the angular position of the peaks) is indicative of (111) fibre texture. Such a texture has often been reported for L1$_0$ FePt films. This apparent texture is sharpest in the upper Pt-rich part of the film, and may explain the absence of certain peaks in these parallel beam X-ray diffraction scans.

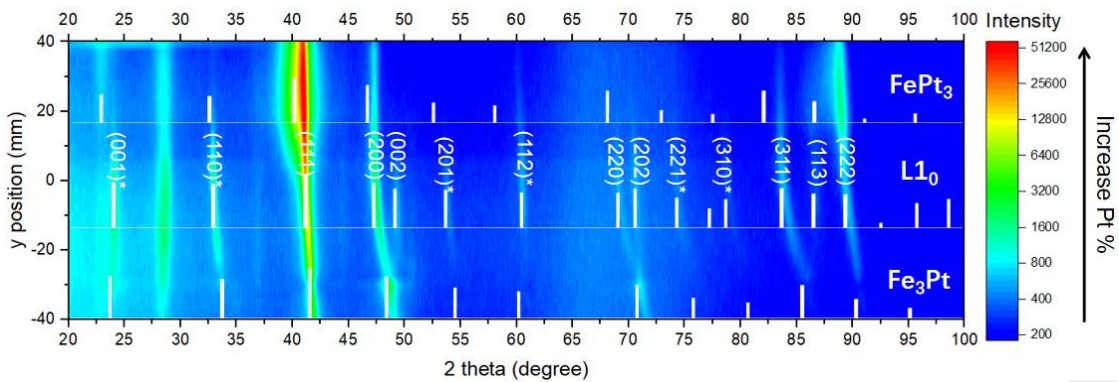

Fig. 8 An XRD waterfall pattern, measured along the vertical section of the film annealed at 600°C for 30 min, identified by the blue rectangle in Figure 7c. The experimental data set is overlaid with the diffraction patterns of chemically ordered L1$_2$-FePt$_3$, L1$_0$-FePt and L1$_2$-Fe$_3$Pt, taken from ICSD reference cards 04-017-4976, 03-065-9121 and 04-001-3171, respectively.

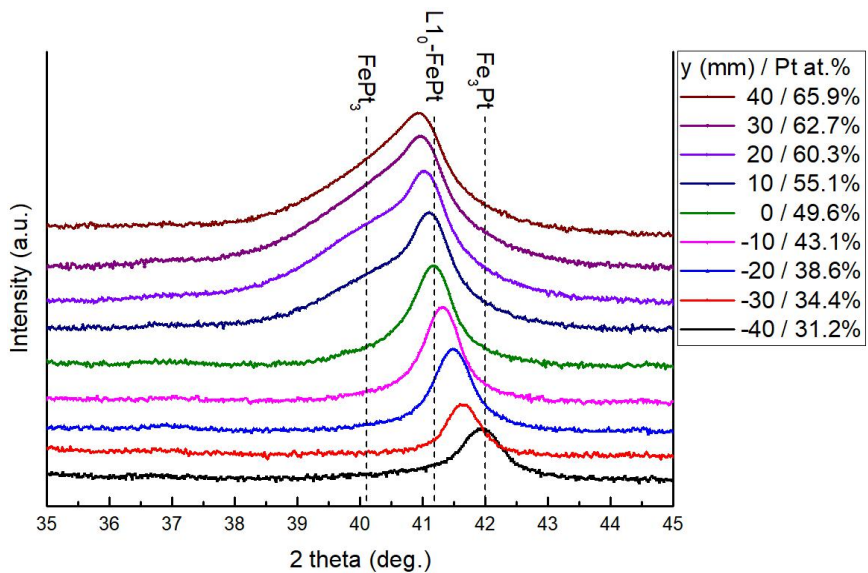

Fig. 9 Zoom of the XRD patterns of the (111) reflection of Figure 8. The experimental data is overlaid with dashed lines indicating the position of the (111) peak of chemically ordered $L1_2$-$FePt_3$, $L1_0$-$FePt$ and $L1_2$-$Fe_3Pt$.

The *a* and *c* lattice parameters, estimated according to the angular position of the (200) and (001) peaks, respectively, are shown as a function of Pt content in Figure 10, and compared with the values of $L1_0$ FePt, ordered and disordered $Fe_3Pt$ and $FePt_3$. Distinct *a* and *c* parameters emerge above 35 at.% Pt, with the *c* value practically reaching that of $L1_0$ FePt above 37.5 at.% Pt, accounting for the emergence of coercivity at this composition. The *a* parameter steadily grows until reaching the value of the $L1_0$ phase at the composition where coercivity is maximum. The lattice parameter of cubic $Fe_{1+x}Pt_{3-x}$, labelled a*, emerges at about 57.5 at.% Pt and then steadily grows with increasing Pt content, indicating improved order with increasing Pt content.

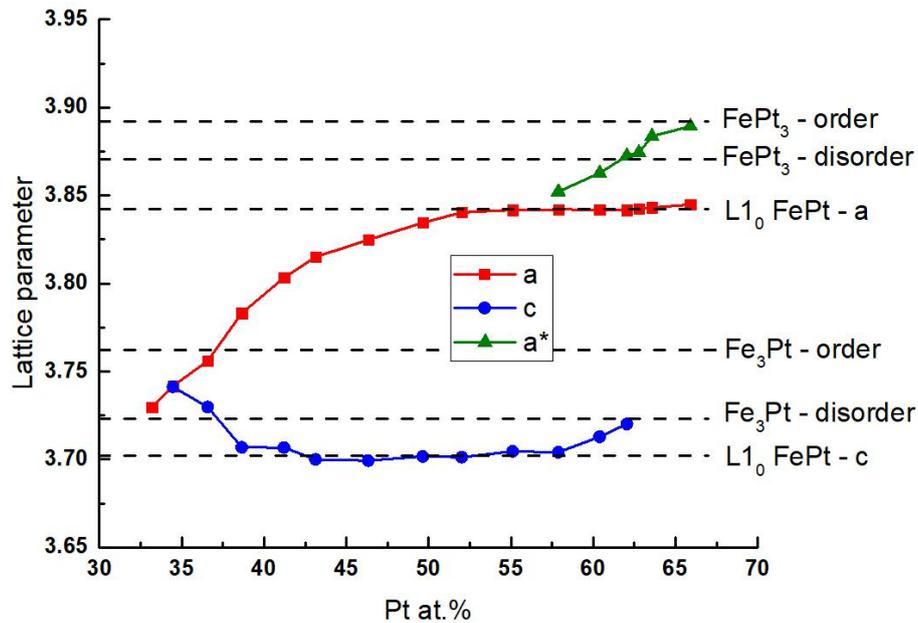

Fig. 10 The *a* and *c* lattice parameters, estimated according to the position of the (200) and (001) peaks, respectively, plotted as a function of Pt content. The a* lattice parameters are estimated according to the position of the lowest angle peak at the Pt-rich end of the film, which corresponds to the (100) peak of the Pt-rich phase. The dashed lines indicate the lattice parameters of $L1_0$-FePt (#03-065-9121), chemically ordered $Fe_3Pt$ (#04-001-3171), chemically disordered $Fe_3Pt$ (#01-071-8366), chemically ordered $FePt_3$ (#04-017-4976) and chemically disordered $FePt_3$(#04-001-8464).

To recap the complementary nature of high throughput characterisation performed on compositionally graded Fe-Pt films, in Figure 11 we replot select sections of XRD patterns together with the film composition, MOKE loops and coercivity values, plotted as a function of position along the vertical section of the film annealed at 600°C for 30 min., identified by a blue rectangular box in Figure 7c. The evolution in the film's coercivity with composition can clearly be correlated with the concomitant evolution in the crystallographic structure, and this compilation of graphs showcases the benefit of high throughput characterisation for screening the properties of compositionally graded hard magnetic films. The emergence of coercivity from the Fe-rich side coincides with the beginning of tetragonal distortion, at the position -25 mm, which has a nominal composition of around 36 at.% Pt. This is in good agreement with what was observed by Ludwig et al. for compositionally graded Fe-Pt films annealed at 500°C and 700°C. However, significant values of coercivity are maintained at higher Pt content in the present study. A difference in the kinetics of phase ordering in both studies, which could account for the difference in magnetic behaviour, may be explained by the fact that in our case the as-deposited film is an alloy, while in the other it is a wedged multilayer stack, requiring interdiffusion of Fe and Pt layers during the annealing process.

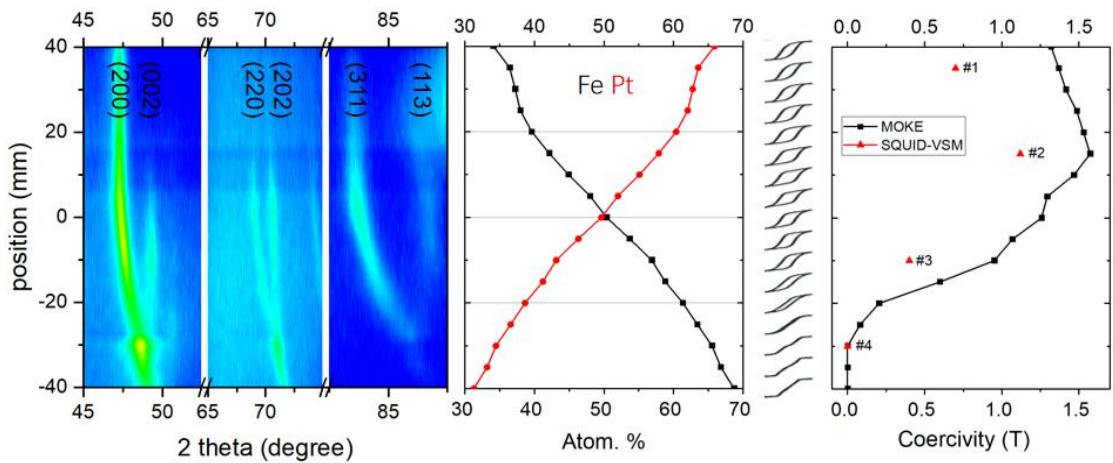

Fig. 11 2D Select sections of XRD patterns together with the film composition from EDX, MOKE loops and coercivity values (black symbols), plotted as a function of position along a vertical section of the film annealed at 600°C for 30 min. Coercivity values estimated from SQUID-VSM measurements of select samples are overlaid on the MOKE coercivity graph (red symbols).

It is important to remark here that the time of acquisition of a 2D set of MOKE loops (on average 323 loops) was approximately 3 h. This is orders of magnitude shorter than the time that would be needed to measure hysteresis loops using a superconducting coil-based magnetometer. Besides, the geometry of the MOKE system allows us to scan over the surface of an intact wafer, while the use of a magnetometer requires the sample to be diced into small pieces with a maximum surface area typically limited to less than 5×5 mm$^2$. Nevertheless, the MOKE technique has a number of limitations. It doesn't measure magnetization but rather a rotation of polarized light which is proportional to magnetization. MOKE measurements probe only the properties at the sample surface, to a depth of a few tens of nm, and are thus not sensitive to any changes in magnetic properties within the volume of the sample. As the technique is based on a measurement of a reflected beam of light, it is not suitable for characterization of rough surfaces. Thus, scanning MOKE is an ideal approach for high throughput characterization to quickly screen the magnetic behavior of thin films, and it can be used to select certain samples for more detailed magnetic characterization using traditional techniques such as SQUID-VSM magnetometry. To validate the MOKE measurements, reference out-of-plane SQUID-VSM measurements were made at room temperature on four pieces of the film annealed at 600°C for 30 minutes (identified by red boxes in Figure 7c). Coercivity values from SQUID-VSM measurements are overlaid on the graph of coercivity values measured by MOKE (Figure 11). The same trend in coercivity as a function of composition is found, though the absolute values are higher for MOKE than for SQUID-VSM (in the samples with highest coercivity, values of 1.13 T and 1.6 T are extracted from SQUID-VSM and MOKE, respectively). The higher values from the MOKE measurements may be related to an over estimate of the applied field value at the surface of the film, and / or to viscosity effects [21, 22] as a very high field sweep rate ($10^6$ T/s) occurs during the MOKE measurement while the SQUID-VSM measurement is made under static field.

*3.2 High throughput preparation and characterization of homogeneous Fe-Pt thin films*

Having demonstrated the use of our scanning MOKE for the high throughput characterization of compositionally graded Fe-Pt films, we will now show another application of it, namely to optimize the fabrication of homogeneous $L1_0$ Fe-Pt films over a large surface area. This is motivated by the desire to be able to integrate such films into parallel processed devices.

A set of films were prepared using targets #4 - #8, which consist of symmetrically positioned pieces of Pt on a base Fe target (Figure 2). The thickness of the FePt layer was reduced to 50 nm, to suit a specific application (high coercivity coatings of Atomic Force Microscope (AFM) probes to be used as Magnetic Force Microscopy (MFM) probes). Based on the study of the influence of annealing temperature and time on compositionally graded films, and to avoid the risk of diffusion during eventual device fabrication, all films were annealed at 600°C for 10 minutes. Figure 12 shows arrays of MOKE loops, together with coercivity maps extracted from these loops, measured from a central section of 50 mm × 50 mm (this area corresponds to the footprint of the AFM probe holder used in a follow-on study). The films made with the minimum (#4) and maximum (#8) amount of Pt show low values of coercivity, with the former having an inhomogeneous coercivity distribution. All other films show rather homogeneous loops and the highest values of coercivity were measured on the film prepared with target #6. An EDX map of a film deposited using this target, shown in Figure 13, indicates that the composition of the film is rather homogeneous. The lower average Pt content (~ 50 at.%) compared to the region of highest coercivity in the compositionally graded films (~ 56 at.%) may reflect a real though unexpected shift in composition or a difference in the estimated value, owing to the difference in thickness of the FePt layer (50 nm in homogeneous films while 100 nm in graded films) that could affect the accuracy of EDX mapping. The red box in Figure 13 indicates the region probed by scanning MOKE, displayed in Figure 12.

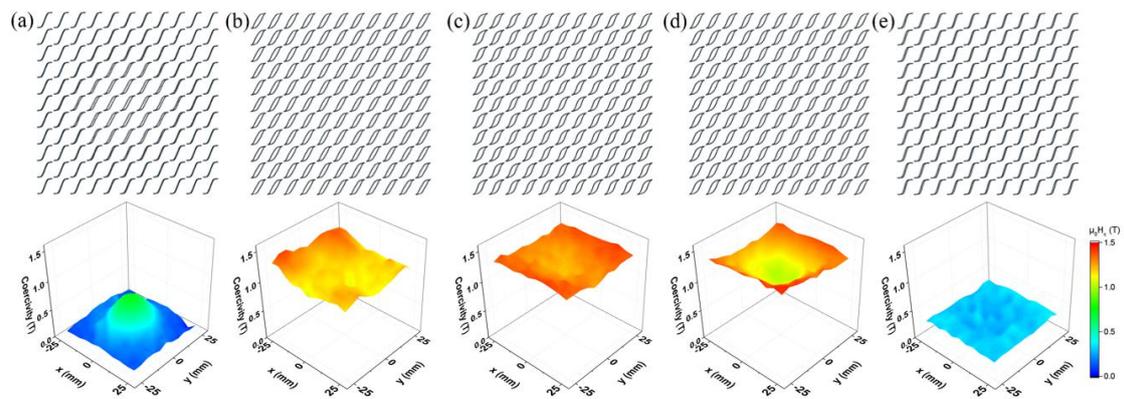

Fig. 12 Polar MOKE loops and coercivity maps on a 50×50 mm² region of films sputtered using targets #4 - #8 and then annealed at 600°C for 10 min.

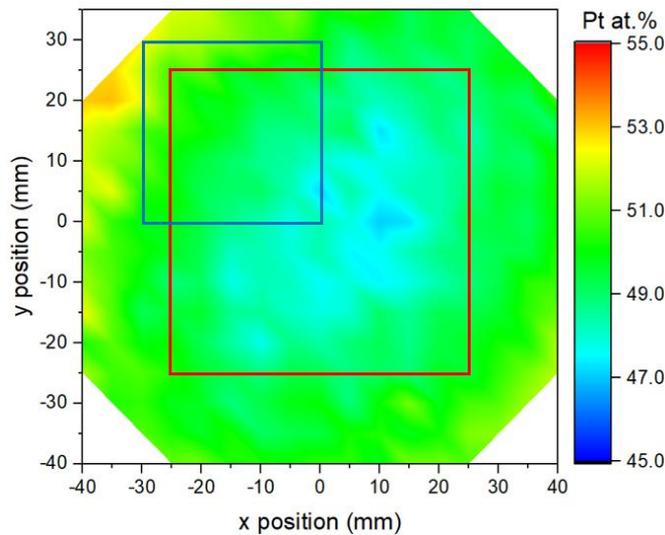

Fig.13 2D EDX map of Fe content measured over a full wafer of a film made using target # 6. The red box indicates the 50×50 mm² wafer zone probed in Figure 12, the blue box the 30×30 mm² region probed in Figure 14.

The RTA furnace used to anneal the entire wafers of compositionally graded and homogeneous films discussed so far, is not necessarily suitable for the annealing of devices containing Fe-Pt films. This is indeed the case for the first targeted application (MFM probes). For this reason, we studied the use of a tube furnace to anneal homogenous Fe-Pt films and scanning MOKE was also used to re-optimise the annealing temperature of films prepared with target #6. In this case, films were diced roughly into quarters, and different quarters were annealed at different temperatures. Figure 14 shows arrays of MOKE loops, together with coercivity maps extracted from these loops, measured from sections of surface area 30 mm × 30 mm (an example of one such area is outlined by the blue box in Figure 13). From this image we can see that slightly higher values of coercivity can be achieved by increasing the nominal annealing temperature from 600°C to 650°C, while annealing at lower or higher temperatures results in a small drop off in coercivity.

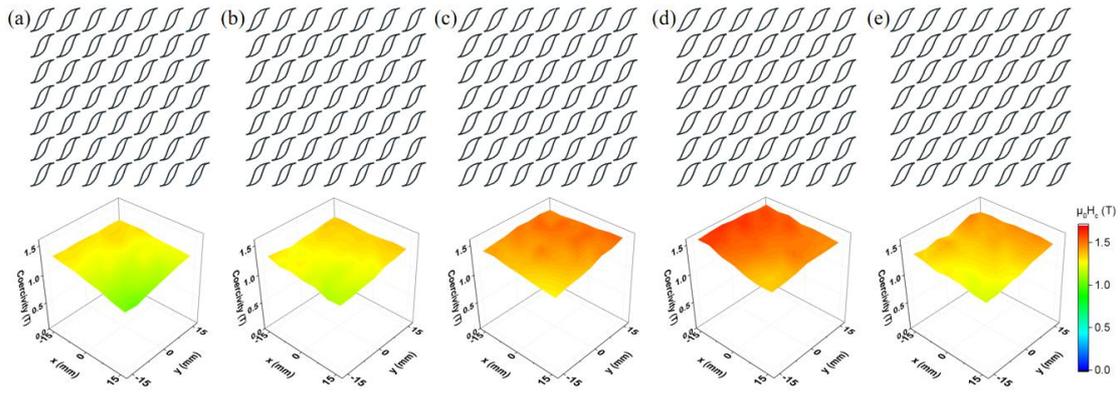

Fig. 14 Hysteresis loops and coercivity maps of sections of films made using target #6, annealed at (a) 500℃, (b) 550℃, (c) 600℃, (d) 650℃, (e) 700℃ for 10 min

4. Conclusions

A set of high throughput scanning characterization techniques (EDX, high field MOKE and XRD) has been used to study compositionally graded 100 nm thick Fe-Pt films. High field MOKE mapping allowed to establish the maximum coercivity values that can be achieved in such Fe-Pt films prepared with the sputtering system used. The triptych of characterization techniques exploited here holds much potential for the high throughput development of hard magnetic materials through the establishment of massive magnetic and structural data sets. In particular, it may be used to optimize the mixing of different rare earths in known Rare Earth Transition Metal phases (2-14-1, 1-5, 1-12), so as to address the RE-balance problem [23, 24]. Beyond this, it could be used to search for new hard magnetic phases in multi-dimensional search spaces. Both challenges will greatly benefit from combining our experimental approach with machine learning approaches, to establish clear correlations between magnetic and structural properties. In a secondary study, the high field MOKE was used to optimize the processing of homogeneous coercive Fe-Pt films, in a step towards their integration into collectively fabricated devices. This demonstrates the tool's potential for use in-line quality control during device production.

5. Acknowledgements

Y.H. received a grant from the China Scholarship Council to carry out part of his PhD training at Institut Néel. This study is partially based on results obtained from the future pioneering program "Development of magnetic material technology for high-efficiency motors (MagHEM), grant number JPNP14015, commissioned by the New Energy and Industrial Technology Development Organization (NEDO), Japan. Funding from the Toyota Motor Corporation is gratefully acknowledged. The LANEF framework (No ANR-10-LABX-51-01) is acknowledged for its support with mutualized infrastructure.